\documentclass[aps,prr,reprint,twocolumn,superscriptaddress,showpacs,floatfix,longbibliography]{revtex4-2}
\usepackage[latin1]{inputenc}
\usepackage{epsfig}
\usepackage{amsfonts}
\usepackage{amssymb}
\usepackage{calrsfs}
\usepackage{amsmath}
\usepackage{xcolor}
\usepackage{graphicx}
\usepackage[colorlinks=true,allcolors=blue]{hyperref}
\hypersetup{breaklinks=true}
\usepackage{lipsum}
\usepackage{bm}
\usepackage[mathscr]{euscript}

\begin{document}
	
	\title{Extended Hydrodynamics of Planar Heisenberg Ferromagnetic Systems}
	\author{Dwigbijoy Paul}
	\email{pauldwigbijoy199218@iisertvm.ac.in}
	\author{Sreedhar B. Dutta}
	\email{sbdutta@iisertvm.ac.in}
	\affiliation{School of Physics, Indian Institute of Science Education and Research Thiruvananthapuram, Thiruvananthapuram 695551, India}
	
	
	\begin{abstract} 
		\noindent 
	
Access to hydro and quasi-hydro regimes in magnetic systems is steadily increasing with the advances in fabrication of clean materials, and motivates study of hydrodynamics beyond conventional regimes and probe non-equilibrium transport. We employ the framework of extended irreversible thermodynamics which includes dynamical dissipative currents as quasi-hydro variables, and effect the modified hydrodynamic equations suitable to describe planar Heisenberg ferromagnetic systems in both paramagnetic and ferromagnetic phases. We show that the new transport coefficients and relaxation time scales introduced by quasi-hydro variables can engineer diverse combination of dispersion relations and provide control over the nature of hydrodynamic modes. For instance, we can change the diffusive energy mode into propagating one in multiple ways, including by tuning the relaxation time of magnetization current.	
	\end{abstract}

	\maketitle
	
	\newcommand{\bee}{\begin{equation}}
		\newcommand{\eee}{\end{equation}}
	\newcommand{\tm}{(t)}
	\newcommand{\gm}{\gamma}
	\newcommand{\dbar}{d\hspace*{-0.08em}\bar{}\hspace*{0.1em}}
	\newcommand{\haf}{\frac{1}{2}}
	\newcommand{\x}{\mathbf{x}}
	\newcommand{\checkk}[1]{\textcolor{red}{#1}}
	
	%
	%
	
	\section{Introduction}\label{intro}
	
	Hydrodynamics is an extensively employed framework for describing universal features of both classical and quantum many-particle systems in the long-wavelength and low-frequency limit. It has been applied to a wide range of systems such as, to name a few, real fluids, spin systems, quark-gluon plasma, electrons in graphene, quasi-particles in Weyl semimetals, fracton phases of quantum matter, and active matter~\cite{Iacocca2019jphyslet,Chaudhuri2013693180,Fritz2024annurevc,Lucas2016pnas1608,Gromov2020physrevr,Marchetti2013revmodph}. Conventional hydrodynamics is effected by identifying the relevant variables based on symmetries and symmetry breaking patterns, and specifying their dynamics adhering to the principles of classical irreversible thermodynamics(CIT). While these coarse-grained dynamical equations can be successfully applied to study the transport phenomena in a multitude of situations~\cite{de2013non,Landau2013}, numerous cases also stipulate appropriate extensions of hydrodynamics~\cite{Brattan2024physrevr,Brewer2015physrevl,Abbasi2024physrevl}. For instance, one of the early motivations to modify Fourier's law and in turn hydrodynamics arose when the resulting heat equation was found to violate causality~\cite{Gavassino2021fspas202}. 
	
	A variety of non-Fourier extensions have been incorporated in a wide range of studies over the decades~\cite{Kovacs2024jphysrep,Cimmelli2010physrevb}. One of the extensively employed systematic extensions of hydrodynamics is based on extended irreversible thermodynamics(EIT), or what essentially amounts to it~\cite{jou1996extended,jou2010thermodynamics,Jou1999201}. EIT is expected to be relevant when the hydrodynamic variables vary rapidly, or their gradients are large, or their relaxation times are long, or when the systems are small, among others. The main tenets of EIT are as follows. First, apart from the classical hydrodynamic variables, the dissipative fluxes are also assumed to be dynamical in nature. Second, the dynamical variables are assumed to be in local non-equilibrium steady state(NESS) instead of being in local equilibrium. Third, NESS is characterized by non-equilibrium entropy that is a function of this extended set of dynamical variables. Finally, the dynamics should be such that the local entropy production rate can never become negative.

	Numerable works with diverse approaches, including statistical-operator frameworks and kinetic theories, were pursued since a long time to provide microscopic justification of the extended hydrodynamics~\cite{luzzi2000nonequilibrium,Vasconcellos1991physreva,Silva2015physreve,Velasco1991physreva}. Predictions based on extended hydrodynamics were also experimentally verified in gaseous systems many decades ago~\cite{Kovacs2019e2107071,Kovacs2021jijengsc}. Yet last decade or so saw a renewed interest in using hydrodynamic frameworks in the study of electron~\cite{Hui2025adfbcd,Narozhny2019jaop2019,Lucas2018aaa274}, phonon~\cite{Guo2015jphysrep,Ghosh2022ac718a} and spin transport~\cite{Flebus2016physrevl,OlguinArias2025zwgxfxkw,Ulloa2019physrevl,Li202290000286,Basso2016physrevb} in solid state systems, primarily due to the advances in fabrication of ultra clean materials that allow hydrodynamic transport. Experimental demonstration of Dirac fluid behavior in graphene~\cite{Crossno2016sciencea,Ku2020s4158602} further led to a flood of hydrodynamic activity including studies beyond classical hydrodynamics~\cite{Tokatly1999physrevb,Tokatly2000physrevb,Lucas2018physrevb}. Observation of thermal transport in various systems, such as suspended graphene~\cite{Lee2015ncomms72}, thin graphite~\cite{Machida2020sciencea} or even organic crystalline polymers~\cite{Zhang2020physrevb}, suggested hydrodynamic phonon transport, further boosting interest in the subject. 
	
	Spin systems have long been fertile ground for not only uncovering a plethora of physical phenomena and advancing a variety of technologies, but also served as toy models for testing various fundamental ideas in low-energy effective theories. Though spin hydrodynamics has been of immense interest for a very long time, it is only in recent experiments the hydrodynamic regime could be accessed in clean ferromagnetic insulators~\cite{Sano2023physrevl}. Magnons are also now realized as spin information carrier in Yttrium-iron garnets(YIG)~\cite{Serga2010264002,Kruglyak2010264001,Flebus2024ad399c}. Collective dynamics of magnons also could be measured recently employing latest experimental techniques~\cite{Xue2026sciencea}. On the theoretical front, studies in hydro~\cite{Iacocca2017physrevl,Iacocca2019physrevb} and quasi-hydro~\cite{Bedell200109500830,Andreev2015s0217979} regimes also indicate renewed interest even in magnetic systems. These experimental and numerical advances that provide access to quasi-hydro regimes of magnetic materials motivate us to investigate extended hydrodynamics of spin systems. 
	
	In this work, we study planar Heisenberg systems with XXZ spin interactions, which are simple model spin systems that can exist in both para-magnetic and ferro-magnetic phases. In Sec.~\ref{Model}, we employ EIT framework to construct explicitly the extended hydrodynamics in both phases. We find, unlike in paramagnetic case, that the dynamical equations in ferro-magnetic phase can contain dissipative current couplings. In Sec.~\ref{Sec_HM}, we analyze linearized extended hydrodynamics in both phases and obtain dispersion relations. We tune various relaxation times scales and current couplings, and investigate the nature of hydrodynamic modes. In Sec.~\ref{conc}, we summarize the current work and discuss feasible application of the formalism developed here to describe other systems.

	%
	%
	\section{Extended irreversible thermodynamics of XXZ spin systems}\label{Model}
	
 The ferromagnetic XXZ Heisenberg spin Hamiltonian describes the planar magnetic order when the exchange interaction in the plane dominates over that along $z$-axis~\cite{HALPERIN1969physrev1,Schwabl1970physrevb,Morozov197803784371,chaikin1995principles}. The Hamiltonian is invariant under time translations and rotations about the $z$-axis. In the high-temperature paramagnetic phase, where the planar magnetization respects the rotational symmetry, the relevant hydrodynamic variables are energy density$~\epsilon(x,t)$ and the $z$-component of magnetization density$~m_z(x,t)$. While in the low-temperature ferromagnetic phase, where the rotational symmetry is broken, the phase$~\theta(x,t)$ of the planar magnetization is also relevant. Unlike in CIT, these quantities form only a partial set of relevant variables in EIT. In the rest of the section, we will specify all the relevant variables and construct their macroscopic dynamics.

	\subsection{Paramagnetic Phase}\label{parahydro}
	
	The symmetries of the XXZ spin Hamiltonian imply that the energy density$~\epsilon(x,t)$ and the $z$-component of magnetization density$~m_z(x,t)$ satisfy the following conservation equations
	\begin{eqnarray}\label{con-eqs}
		\frac{\partial \epsilon}{\partial t} = - \nabla \cdot J_{\epsilon} \quad , \quad \frac{\partial m_z}{\partial t} = - \nabla \cdot J_m ~,
	\end{eqnarray}   
	where $J_{\epsilon} = J_{\epsilon}(x,t)$ and $J_m =J_m(x,t)$ are the current densities of energy and magnetization, respectively. 
	
	In hydrodynamics based on CIT, the current densities are assumed to be non-dynamical quantities satisfying constitutive relations, while the dynamics of thermodynamic variables is hypothesized to respect local equilibrium. Hence, by employing the Gibbs relation for equilibrium states 
	\begin{equation}\label{Gibbs}
		Tds = d\epsilon - h dm_z~,
	\end{equation}
	  where $s,T$ and $h$ denote entropy density, temperature and magnetic field, respectively, and using the conservation equations, we can deduce the entropy balance equation 
	  \begin{equation}\label{ent-eq}
	  	\frac{\partial s}{\partial t} + \nabla \cdot J_s = \sigma~,
	  \end{equation}
	 where the entropy flux$~J_s$ and the entropy production$~\sigma$ are given by the expressions
	 \begin{eqnarray}\label{ent-cur-prod}
	 	J_s = \frac{Q}{T} \quad,\quad Q= J_{\epsilon} - h J_m , \nonumber\\
	 	\sigma = -\frac{1}{T^2} Q \cdot \nabla T -\frac{1}{T} J_m \cdot \nabla h~,
	 \end{eqnarray}
	where$~Q$ denotes heat flux and is often traded for$~J_{\epsilon}$. The constitutive relations are then established perturbatively in gradient expansion by assuming that the local entropy production is non-negative, namely $\sigma \ge 0$. The explicit expressions of the dissipative currents to leading order are
	\begin{eqnarray}\label{cons-rel-1}
		Q =  - \kappa_{11} \nabla T- \kappa_{12} \nabla h ~,\nonumber \\
		J_m = -\kappa_{21} \nabla T- \kappa_{22} \nabla h ~,
	\end{eqnarray} 
	where the coefficients$~\kappa_{ij}$ in general can depend on equilibrium state variables.  
	
	The local equilibrium hypothesis essentially implies neglecting local relaxation time scales compared to hydrodynamic time scales, and assuming uniform surroundings for a {\it fluid element}. But, more the variation of hydrodynamic variables in space and time, less is the validity of these assumptions. More generally, a {\it fluid element} is surrounded by an environment with varying temperature$~T$ and magnetic field$~h$, and is more appropriate to consider the element to be in a steady state rather than in equilibrium. Thus, in the hydrodynamics based on EIT, we assume that the dynamics is consistent with local NESS instead of local equilibrium. We also relax the assumption that relaxation time scales of currents are negligible. 
	
	The conventional equilibrium thermodynamic variables are no longer sufficient and additional non-equilibrium variables are required to characterize NESS.  Non-equilibrium extension of thermodynamics is assumed to be given by an extended fundamental relation  
	\begin{equation}\label{mod-fun}
		s= s(\epsilon, m_z, J_{\epsilon}, J_m)~,
	\end{equation}  
	where the entropy also depends on dissipative currents apart from the equilibrium thermodynamic variables$~\epsilon$ and$~m_z$. This leads to a modified Gibbs relation
	\begin{equation}\label{mod-gibbs}
		Tds= d\epsilon - h dm_z -\alpha_{\epsilon}\cdot dJ_{\epsilon} - \alpha_{m} \cdot dJ_m~,
	\end{equation}
	that further defines two additional conjugate vectors
	\begin{eqnarray}\label{con-vecs}
		\alpha_{\epsilon} = -T\frac{\partial s}{\partial J_{\epsilon}} ~,~ \alpha_{m} = -T \frac{\partial s}{\partial J_{m}}~,
	\end{eqnarray} 
	which can be written in the form
	\begin{eqnarray}\label{con-vecs-form}
		\alpha_{\epsilon}  =  \alpha_{11} J_{\epsilon} +  \alpha_{12} J_{m}~,\quad
	\alpha_{m}  =  \alpha_{21} J_{\epsilon} +  \alpha_{22} J_{m}~,
	\end{eqnarray}
	where $\alpha_{ij}$ are scalar functions of state variables that are constrained by tensor identities$~\partial{\alpha_{\epsilon} }/\partial{J_{m}}  = \partial{\alpha_{m} } / \partial{J_{\epsilon}}$.
	
	 Essentially, in the extended hydrodynamic description the currents$~J_{\epsilon}$ and$~J_m$ are also elevated to dynamical fields, and can possess nontrivial dynamics since it is only required to be consistent with the modified Gibbs relation. We construct these dynamical equations by determining the expression for entropy production in this modified set up and then demanding it to be necessarily non-negative. Up on using the conservation equations(\ref{con-eqs}) of energy and magnetization even in the modified Gibbs relation(\ref{mod-gibbs}), we again arrive at the balance equation as given in Eq.(\ref{ent-eq}). Suppose we choose the entropy flux $J_{s}= Q/T$, same as in the CIT case given by Eq.(\ref{ent-cur-prod}), then the entropy production takes the modified form given by
	\begin{eqnarray}
		\sigma &= &-\frac{1}{T} J_{\epsilon} \cdot \left(  \alpha_{11}\frac{\partial J_{\epsilon}}{\partial t} +  \alpha_{21}\frac{\partial J_{m}}{\partial t} -T \nabla \frac{1}{T}  \right) \nonumber \\
		 &-&\frac{1}{T} J_{m} \cdot \left(\alpha_{12}\frac{\partial J_{\epsilon}}{\partial t} +   \alpha_{22}\frac{\partial J_{m}}{\partial t} +  T\nabla \frac{h}{T} \right).
	\end{eqnarray}
	Hence, imposing the constraint that$~\sigma \ge 0$ has to necessarily hold for any physically allowed solution of the modified hydrodynamics leads to the following current equations,
	\begin{eqnarray}\label{curr-eqn-ver1}
		\alpha_{11}\frac{\partial J_{\epsilon}}{\partial t} &+&  \alpha_{21}\frac{\partial J_{m}}{\partial t} -T \nabla \frac{1}{T} = -A_{11} J_{\epsilon}  - A_{12} J_{m},\nonumber \\
		\alpha_{12}\frac{\partial J_{\epsilon}}{\partial t} &+&   \alpha_{22}\frac{\partial J_{m}}{\partial t} +  T\nabla \frac{h}{T} = -A_{21} J_{\epsilon} - A_{22}  J_{m},
	\end{eqnarray} 
	where$~A_{ij}$ are NESS dependent variables. Essentially, under the dynamics governed by Eqs.(\ref{con-eqs}) and Eqs.(\ref{curr-eqn-ver1}) the entropy production has the form
	\begin{eqnarray}
		\sigma = \frac{1}{T} 
			\begin{pmatrix}
			J_{\epsilon} &
			J_{m}
		\end{pmatrix}
		\begin{pmatrix}
			A_{11} & A_{12} \\
			A_{21} & A_{22}
		\end{pmatrix}
		\begin{pmatrix}
			J_{\epsilon} \\
			J_{m}
		\end{pmatrix},
	\end{eqnarray}
	in the matrix notation. The case$~\sigma=0$ corresponds to adiabatic flows where the currents necessarily vanish, while the case$~\sigma >0$ is associated with dissipative flows with non-vanishing currents. Therefore, the matrix$~A$ is constrained to be positive definite. It should be remarked that the choice of entropy flux is in fact the most general form, though we could have added a vector quantity$~{c_{\epsilon} J_{\epsilon} + c_m J_m}$, where$~c_{\epsilon}$ and$~c_m$ are state dependent scalar quantities. It then turns out that this would lead to additional terms in$~\sigma$ which violate positivity condition unless$~c_{\epsilon}$ and$~c_m$ vanish.
	
	Furthermore, since the current equations(\ref{curr-eqn-ver1}) are expected to reduce to the constitutive relations(\ref{cons-rel-1}) in the large time limit, the variables$~A_{ij}$ can be reparametrized in terms of$~\kappa_{ij}$ which are extended transport coefficients that depend on NESS.  For later convenience, we write down the explicit expression of the inverse matrix 
	\begin{eqnarray}\label{A-inv}
		A^{-1} =
		\begin{pmatrix}
			T(\kappa_{11} \!+ \!h \kappa_{21}) \!+\! h(\kappa_{12}\!+\! h \kappa_{22}) & \kappa_{12}\!+\! h \kappa_{22}\\
				T\kappa_{21}\!+\! h \kappa_{22} & \kappa_{22}
		\end{pmatrix}.
	\end{eqnarray}
Now the current equations can be rewritten as
\begin{widetext}
	\begin{eqnarray}\label{curr-eqn-para-eit}
			\tau_{11}\frac{\partial J_{\epsilon}}{\partial t} +  \tau_{12}\frac{\partial J_{m}}{\partial t} +  J_{\epsilon} &=& -(\kappa_{11} + h \kappa_{21}) \nabla T
			 - (\kappa_{12} + h \kappa_{22})  \nabla h~,\nonumber \\
		\tau_{21}\frac{\partial J_{\epsilon}}{\partial t} +   \tau_{22}\frac{\partial J_{m}}{\partial t} + J_{m} &=& -\kappa_{21} \nabla T- \kappa_{22} \nabla h ~,	
	\end{eqnarray} 
\end{widetext}
	where the parameters$~\tau_{ij}$ are defined in the matrix relation 
	\begin{eqnarray}\label{tau-mat}
			\begin{pmatrix}
			\tau_{11} & \tau_{12} \\
			\tau_{21} & \tau_{22}
		\end{pmatrix}= A^{-1} 	\begin{pmatrix}
		\alpha_{11} & \alpha_{21} \\
		\alpha_{12} & \alpha_{22}
		\end{pmatrix}.
	\end{eqnarray}
	The above $\tau$-matrix should be positive definite to ensure that the current equations(\ref{curr-eqn-para-eit})  asymptotically relax to constitutive relations. In the absence of cross coefficients$~\tau_{12}$ and$~\tau_{21}$, the parameters$~\tau_{11}$ and$~\tau_{22}$ are interpreted as timescale for $J_{\epsilon}$ and$~J_{m}$ to become non-dynamical, respectively. The cross coefficients essentially couple these two relaxation timescales, thus providing, for instance, a means of controlling magnetic profiles with heat currents or regulating heat dissipation with magnetic fields. 
	
	To sum up, the modified hydrodynamics in paramagnetic phase is described by Eqs.(\ref{con-eqs}) and~(\ref{curr-eqn-para-eit}), along with the modified fundamental relation(\ref{mod-fun}) or equivalently the corresponding equations of state, where $T,h$ and$~\alpha_{ij}$ are given functions of NESS variables. The system of equations is complete when either transport coefficients$~\kappa_{ij}$ or relaxation time parameters$~\tau_{ij}$ are further specified in terms of NESS variables.

	\subsection{Ferromagnetic Phase}\label{ferrohydro}
	
	At low temperatures, the planar magnetization acquires a nonzero expectation value and breaks the rotational symmetry. As we know, this leads to the emerge of another relevant variable in the classical hydrodynamics, namely the phase~$\theta(x,t)$ of the quantity$~{m_x+im_y}$, or rather the deviation of phase from its expectation value. The thermodynamic state is characterized by$~\epsilon, m_z$ and a curl-free vector$~v_{\theta}=\nabla \theta$. In the ferromagnetic phase, the Gibbs relation  is 
		\begin{equation}\label{Gibbs-ferro}
		Tds = d\epsilon - h dm_z - h_{\theta} \cdot dv_{\theta} ~,
	\end{equation}
	where the vector quantity
	\begin{equation}\label{h-theta-def}
		h_{\theta} = -T\frac{\partial s}{\partial v_{\theta}},
	\end{equation}
	is thermodynamic conjugate of$~v_{\theta}$. The corresponding hydrodynamics, apart from Eqs.(\ref{con-eqs}), also includes a balance equation$~	\partial_t \theta = h - \psi$,
	or equivalently,
		\begin{equation}\label{phase-eqn}
		\frac{\partial v_{\theta}}{\partial t} = \nabla h - \nabla \psi~,
	\end{equation}
	where$~\psi$ denotes the dissipative current associated with the phase. The dynamical equations of$~\epsilon, m_z$ and$~v_{\theta}$ and the local equilibrium hypothesis of CIT leads to the entropy equation(\ref{ent-eq}), where the entropy flux
		 \begin{eqnarray}\label{ent-flux-ferro-cit}
		J_s = \frac{Q}{T} \quad,\quad Q= J_{\epsilon} - h J_m - h_{\theta} \psi ,
	\end{eqnarray}
and the entropy production
\begin{equation}\label{ent-cur-prod-ferro}
	\sigma = \left(J_{\epsilon}-h J_m \right)\cdot \nabla \frac{1}{T} -\frac{1}{T}\left(J_m + h_{\theta} \right)\cdot \nabla h - \psi \nabla \cdot \frac{h_{\theta}}{T}.
\end{equation}
In absence of dissipation, entropy production vanishes identically and hence$~{J_{\epsilon}=-h h_{\theta}}, {J_m =- h_{\theta}}$, and$~{\psi=0}$. The dissipative effects modify these currents given by  
	\begin{eqnarray}\label{cons-rel-2}
	J_{\epsilon}&=& -h h_{\theta} -\kappa_{11} \nabla T- \kappa_{12} \nabla h ,\nonumber \\
	J_m &=& -h_{\theta}-\kappa_{21} \nabla T- \kappa_{22} \nabla h , \nonumber \\ 
	 \psi &=& - \gamma \nabla \cdot \frac{h_{\theta}}{T}.
\end{eqnarray}
The positivity of$~\sigma$ in presence of dissipation implies that$~\kappa_{ij}$ has the same properties as mentioned before and the parameter$~{\gamma>0}$. The isotropic transport coefficients$~\kappa_{ij}$ considered here for simplicity are not the most general choice in the symmetry broken phase. 

The extended hydrodynamics based on EIT in the ferromagnetic phase further includes the quantity$~\psi$ as a dynamical variable. It is convenient to separate the dissipative part of the currents and write the equations of motion for$~\epsilon$ and$~m_z$ as
	\begin{equation}\label{con-eqs-ferro}
	\frac{\partial \epsilon}{\partial t} =  \nabla \cdot \left( h h_{\theta} \right) - \nabla \cdot J_{\epsilon},~ \frac{\partial m_z}{\partial t} = \nabla \cdot  h_{\theta}  - \nabla \cdot J_m .
\end{equation}
NESS is assumed to be characterized by a modified fundamental relation  
\begin{equation}\label{mod-fun-ferro}
	s= s(\epsilon, m_z, v_{\theta}, J_{\epsilon}, J_m, \psi)~,
\end{equation}  
leading to a modified Gibbs relation
\begin{equation}\label{mod-gibbs-ferro}
	Tds= d\epsilon - h dm_z  - h_{\theta} \cdot dv_{\theta}-\alpha_{\epsilon}\cdot dJ_{\epsilon} - \alpha_{m} \cdot dJ_m -\alpha_{\theta}d\psi~,
\end{equation}
where the scalar variable$~\alpha_{\theta}$ is thermodynamic conjugate of quantity$~\psi$. If we assume that the conjugate variables$~\alpha_{\epsilon}, \alpha_{m}$ and$~\alpha_{\theta}$ vanish when dissipative currents are turned off, then based on the tensorial properties these variables in general can be written as 
	\begin{eqnarray}\label{h-theta-def}
		\alpha_{\epsilon} &=& -T\frac{\partial s}{\partial J_{\epsilon}} = \alpha_{11} J_{\epsilon} +  \alpha_{12} J_{m} + \beta_1 \psi v_{\theta}, \nonumber \\
		\alpha_{m} &=& -T \frac{\partial s}{\partial J_{m}}  = \alpha_{21} J_{\epsilon} +  \alpha_{22} J_{m} + \beta_2 \psi v_{\theta},\nonumber \\
	\alpha_{\theta} &=& -T \frac{\partial s}{\partial \psi} = \beta_3 \psi ,
\end{eqnarray}
where the four $\alpha$-parameters and the three $\beta$-parameters are scalar  functions of NESS variables. We can obtain the entropy balance equation using Eqs.(\ref{phase-eqn}),(\ref{con-eqs-ferro}) and(\ref{mod-gibbs-ferro}). The entropy flux has a modified form
\begin{equation}\label{ent-flux-ferro}
	J_s = \frac{Q}{T} + \beta_{s1} \psi J_{\epsilon} + \beta_{s2} \psi J_m,
\end{equation}
where$~\beta_{s1}$ and$~\beta_{s2}$ are scalar functions of the state variables. In this case, the positivity of entropy production can be maintained even with non-zero $~\beta_{s1}$ and$~\beta_{s2}$, which as we will see appear in the dynamical equations. The variables$~\beta_{s1}$ and$~\beta_{s2}$ can be interpreted as coupling of phase current with energy and magnetization currents, respectively. The corresponding entropy production is found to be 
\begin{widetext}
\begin{eqnarray}\label{ent-prod-ferro}
		\sigma =  &-&\frac{J_{\epsilon}}{T} \cdot \left[  \alpha_{11}\frac{\partial J_{\epsilon}}{\partial t} +  \alpha_{21}\frac{\partial J_{m}}{\partial t}  -T\nabla \left( \frac{1}{T} + \beta_{s1} \psi \right) \right] 
	-\frac{J_{m}}{T}  \cdot \left[ \alpha_{12}\frac{\partial J_{\epsilon}}{\partial t} +   \alpha_{22}\frac{\partial J_{m}}{\partial t} + T\nabla \left(\frac{h}{T} - \beta_{s2} \psi \right) \right] \nonumber \\
	&-& \frac{\psi}{T} \left[\beta_3 \frac{\partial \psi}{\partial t}
	+ v_{\theta} \cdot \left(\beta_1 \frac{\partial J_{\epsilon}}{\partial t} + \beta_2 \frac{\partial J_{m}}{\partial t}\right) + T \left(\nabla \cdot\frac{h_{\theta}}{T} - \beta_{s1} \nabla \cdot J_{\epsilon} - \beta_{s2} \nabla \cdot J_m \right)  \right].
\end{eqnarray}
\end{widetext}
As expected, $~\sigma=0$ when the dissipative currents vanish. Furthermore, imposing the positivity condition on$~\sigma$ leads to the following dynamical equations
\begin{widetext}
\begin{eqnarray}\label{curr-eqn-ferro-ver1}
	&&\alpha_{11}\frac{\partial J_{\epsilon}}{\partial t} +  \alpha_{21}\frac{\partial J_{m}}{\partial t}  -T\nabla \left( \frac{1}{T} + \beta_{s1} \psi \right) =  -A_{11} J_{\epsilon}  - A_{12} J_{m},\nonumber \\
	&&\alpha_{12}\frac{\partial J_{\epsilon}}{\partial t} +   \alpha_{22}\frac{\partial J_{m}}{\partial t} + T\nabla \left(\frac{h}{T} - \beta_{s2} \psi \right) =  -A_{21} J_{\epsilon}  - A_{22} J_{m},\nonumber \\
	&&\beta_3 \frac{\partial \psi}{\partial t}
	+ v_{\theta} \cdot \left(\beta_1 \frac{\partial J_{\epsilon}}{\partial t} + \beta_2 \frac{\partial J_{m}}{\partial t}\right) + T \left(\nabla \cdot\frac{h_{\theta}}{T} - \beta_{s1} \nabla \cdot J_{\epsilon} - \beta_{s2} \nabla \cdot J_m \right) = -\lambda \psi,
\end{eqnarray}
\end{widetext}
where$~\lambda$ is another transport coefficient. When the time and spatial derivatives of the dissipative currents are neglected, along with gradients of$~\beta_{s1}$ and$~\beta_{s2}$, the above dynamical equations reduce to the constitutive relations(\ref{cons-rel-2}). The familiar form of constitutive relations manifest when we rewrite Eqs.(\ref{curr-eqn-ferro-ver1}) as
\begin{widetext}
	\begin{eqnarray}\label{curr-eqn-ferro-eit}
		&&\tau_{11}\frac{\partial J_{\epsilon}}{\partial t} +  \tau_{12}\frac{\partial J_{m}}{\partial t} +  J_{\epsilon} = (A^{-1})_{11} T\nabla \left( \frac{1}{T} + \beta_{s1} \psi \right)  - (A^{-1})_{12}  T\nabla \left(\frac{h}{T} - \beta_{s2} \psi \right),\nonumber \\
		&&\tau_{21}\frac{\partial J_{\epsilon}}{\partial t} +   \tau_{22}\frac{\partial J_{m}}{\partial t} + J_{m} = (A^{-1})_{21} T\nabla \left( \frac{1}{T} + \beta_{s1} \psi \right) - (A^{-1})_{22}  T\nabla \left(\frac{h}{T} - \beta_{s2} \psi \right), \nonumber \\
		&&\tau_3 \frac{\partial \psi}{\partial t} + \psi
		+ v_{\theta} \cdot \left(\tilde{\beta}_1 \frac{\partial J_{\epsilon}}{\partial t} + \tilde{\beta}_2 \frac{\partial J_{m}}{\partial t}\right) =- \gamma \left(\nabla \cdot\frac{h_{\theta}}{T} - \beta_{s1} \nabla \cdot J_{\epsilon} - \beta_{s2} \nabla \cdot J_m \right),
	\end{eqnarray} 
\end{widetext}
where we have relabeled various quantities that appear in Eq.(\ref{curr-eqn-ferro-ver1}), using equations(\ref{A-inv}),(\ref{tau-mat}), ${\lambda=T/\gamma}$, ${\beta_1=T \tilde{\beta}_1/\gamma}$, ${\beta_2=T \tilde{\beta}_2/\gamma}$, and$~{\beta_3=T\tau_3/\gamma}$. We see that the dynamics of energy and magnetization is further coupled by the presence of phase current, and thus can provide additional handle in controlling, say, heat dissipation with magnetic fields. 

To summarize, the extended hydrodynamics in ferromagnetic phase is captured by Eqs.(\ref{phase-eqn}), (\ref{con-eqs-ferro}), and~(\ref{curr-eqn-ferro-eit}), along with the extended fundamental relation(\ref{mod-fun-ferro}) and Gibbs relation(\ref{mod-gibbs-ferro}). The system of equations is complete up on further specifying the dependence of relaxation coefficients$~\tau_{ij}$s and$~\tau_3$, and the kinetic coefficients $~\beta_{s1}, \beta_{s2}$ and$~\gamma$ on the steady state variables.

	%
	%
	
	\section{Hydrodynamic modes}\label{Sec_HM}
	
	In this section, we analyze the propagating and diffusive nature of hydrodynamic modes in both para- and ferro-magnetic phases by means of dispersion relations obtained from the corresponding linearized hydrodynamics. We will see that the quasi-hydrodynamic modes associated with dissipative currents can collude with classical hydrodynamic modes resulting in collective modes of entirely different nature.

	\subsection{Dispersion relations in Paramagnetic Phase}\label{Para-modes}
	
	The dispersion relations of classical hydrodynamic modes are of course well know. Linearizing Eqs.(\ref{con-eqs}) and~(\ref{cons-rel-1}) about equilibrium values result in equations of the form
	\begin{eqnarray}\label{lin-para-cl}
			\frac{\partial \epsilon}{\partial t} &=& \bar{\kappa}_{11}c_m^{-1}\nabla^2 \epsilon + \bar{\kappa}_{12}\chi^{-1} \nabla^2 m_z ,
		\nonumber \\ \frac{\partial m_z}{\partial t} &=& \kappa_{21}c_m^{-1} \nabla^2 \epsilon + \kappa_{22}\chi^{-1} \nabla^2 m_z,
	\end{eqnarray}
	where $\bar{\kappa}_{11} = \kappa_{11} + \kappa_{12}\chi^{-1}m_{z0}$, $\bar{\kappa}_{12} = \kappa_{12} + \kappa_{22}\chi^{-1}m_{z0}$, and the parameters$~c_m, \chi$ and$~m_{z0}$ denote specific heat, susceptibility and initial average magnetization, respectively. In obtaining Eq.(\ref{lin-para-cl}), we have used the linear relations ${T=c_m^{-1}\epsilon}$ and ${h=\chi^{-1}}m_z$, neglected nonlinear terms such as$~(\partial m_z)^2, (\partial T)^2$ and$~\partial T \partial m_z$, and employed the same notation to denote deviations from the equilibrium state. It should be remarked that it is more common to linearize about equilibrium states with$~T=T_0$ and$~m_z=0$ in the paramagnetic phase. In this case, the cross-coefficients$~\kappa_{12}$ and$~\kappa_{21}$ vanish since they are odd variables under time-reversal and necessarily of$~O(m_{z0})$. Thus the energy and magnetization decouple from each other, following the dispersion relations$~{\omega= -i D_{\epsilon} k^2}$  and$~{\omega= -i D_m k^2}$, respectively, where the diffusion coefficients
	\begin{equation}\label{DeDm-def}
	D_{\epsilon}= \kappa_{11} / c_m ~, \quad  D_m= \kappa_{22} /\chi~.
	\end{equation}
	 However, since the $z$-component of total magnetization is conserved, $m_z$ asymptotically will approach a non-zero value$~m_{z0}$  for any non-zero initial magnetization. Hence, in general, non-vanishing cross-coefficients can be chosen. Though the two modes couple, they remain diffusive and follow the dispersion$~{\omega_{\pm} = -i D_{\pm} k^2}$, where 
\begin{equation}\label{diff-coeff}
	D_{\pm}= \frac{1}{2} \left( \frac{\bar{\kappa}_{11}}{c_m} +  \frac{\kappa_{22}}{\chi} \right) \pm \sqrt{ \frac{1}{4}\left( \frac{\bar{\kappa}_{11}}{c_m} -  \frac{\kappa_{22}}{\chi} \right)^2  +   \frac{\bar{\kappa}_{12}\kappa_{21}}{c_m\chi} }.
\end{equation}
These diffusion coefficients are ensured to be positive due to the stability condition$~{ \bar{\kappa}_{11} \kappa_{22} > \bar{\kappa}_{12}\kappa_{21} }$.

The linearization of extended hydrodynamical equations(\ref{curr-eqn-para-eit}) lead to 
	\begin{eqnarray}\label{lin-para-eit}
		\tau_{11}\frac{\partial J_{\epsilon}}{\partial t} +  \tau_{12}\frac{\partial J_{m}}{\partial t} +  J_{\epsilon} &=& -\frac{\bar{\kappa}_{11}}{c_m}\nabla \epsilon - \frac{\bar{\kappa}_{12}}{\chi} \nabla m_z,
		\nonumber \\
		\tau_{21}\frac{\partial J_{\epsilon}}{\partial t} +   \tau_{22}\frac{\partial J_{m}}{\partial t} + J_{m} &=& -\frac{\kappa_{21}}{c_m} \nabla \epsilon - \frac{\kappa_{22}}{\chi} \nabla m_z,
	\end{eqnarray} 
	where$~\tau_{11}$ and$~\tau_{22}$ are of$~O(1)$, while$~\tau_{12}$ and$~\tau_{21}$ are of$~O(m_{z0})$. These equations imply that the transverse components of energy and magnetization currents couple due to cross coefficients$~\tau_{12}$ and$~\tau_{21}$, and decay with time scales
	\begin{equation}\label{relax-times}
		\tau_{\pm}= \frac{1}{2} \left( \tau_{11} +\tau_{22} \right) \pm \sqrt{ \frac{1}{4}\left( \tau_{11}-\tau_{22} \right)^2  +  \tau_{12}\tau_{21} }.
	\end{equation}
			\begin{figure}[t]
		\centering
		\includegraphics[width=\columnwidth]{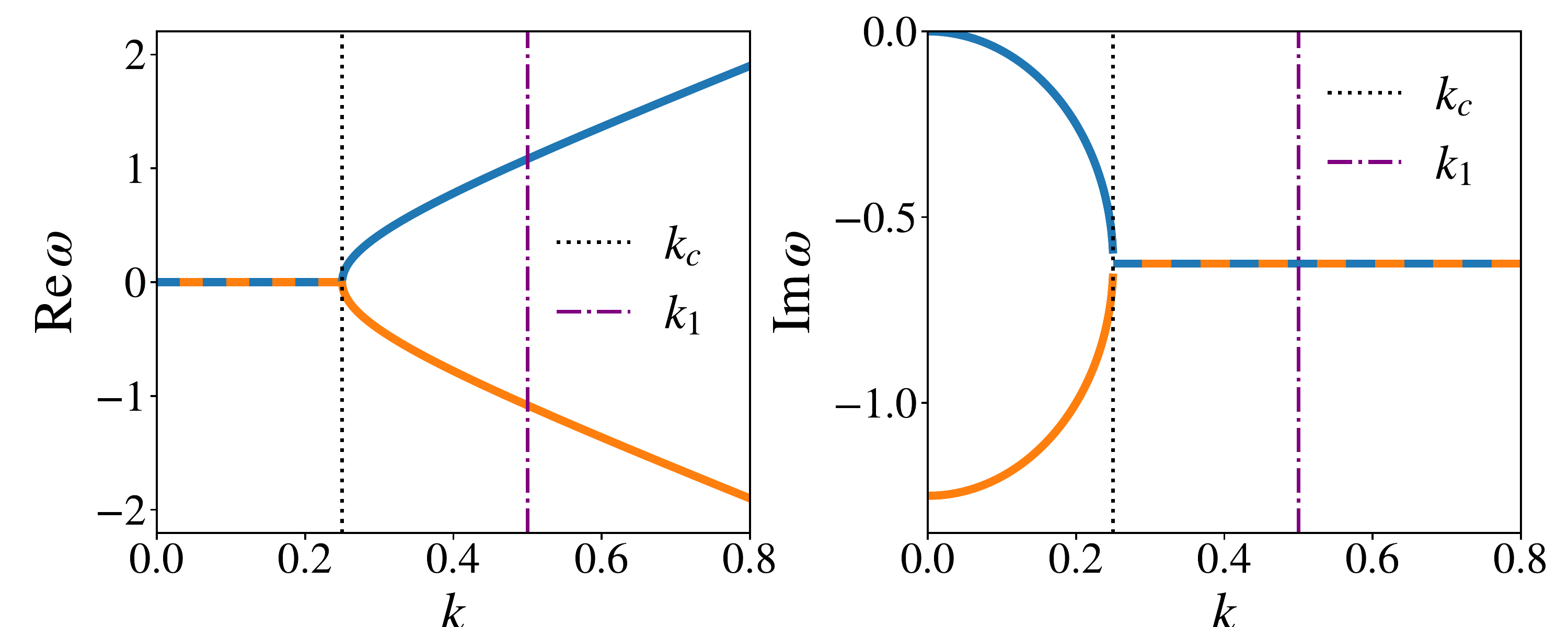}
		\caption{Real and imaginary parts of EIT dispersion relations of energy sector in paramagnetic phase for$~{\tau_{11}=0.8}$ and$~{D_{\epsilon}=5}$. Cross-over from diffusive to propagating modes occur at$~{k_c=k_1/2}$.}\label{fig1}
	\end{figure}
	While the longitudinal components couple with energy and magnetization densities resulting in the Fourier-Laplace transformed equations
	\begin{widetext}
		\begin{eqnarray}\label{para-eit-four}
				\begin{bmatrix}
		-i\omega & i|k| & 0 & 0 \\
		i\bar{\kappa}_{11} c_m^{-1} |k|& 1-i\omega \tau_{11} & i\bar{\kappa}_{12} \chi^{-1} |k|& -i\omega \tau_{12}\\
		0 & 0 & -i\omega  &  i|k| \\
	   i\kappa_{21} c_m^{-1}|k|	& -i\omega \tau_{21}& i \kappa_{22} \chi^{-1}|k| &  1-i\omega \tau_{22}
	\end{bmatrix}
		\begin{bmatrix}
		\tilde{\epsilon} (k,\omega)\\
		 \tilde{J}^l_{\epsilon}(k,\omega)\\
		 \tilde{m}_z(k,\omega)\\
		 \tilde{J}^l_m(k,\omega)
	\end{bmatrix}
	=
	\begin{bmatrix}
	1	& 0 & 0 & 0 \\
	0	&  \tau_{11} & 0 & \tau_{12} \\
	0	&  0 & 1 & 0\\
	0	& \tau_{21} &0 & \tau_{22} 
	\end{bmatrix}
			\begin{bmatrix}
		\epsilon (k, 0)\\
		J^l_{\epsilon}(k,0)\\
		m_z(k,0)\\
		J^l_m(k,0)
	\end{bmatrix},
		\end{eqnarray} 
	\end{widetext}
	where$~\tilde{f}(k,\omega) = \int d^3x \int_{0}^{\infty} dt  \exp(i(\omega t- k \cdot x) ) f(x,t)$ denotes Fourier-Laplace transform of a function$~f(x,t)$, while$~f(k,0)=\int d^3x \exp(-i k \cdot x) {f(x,{t=0})}$ denotes Fourier transform of initial function$~{f(x,{t=0})}$. The determinant of the matrix appearing on left-hand side of Eq.(\ref{para-eit-four}) is the polynomial
	\begin{equation}\label{det-para}
	  P_4(\omega, k)= P_{2\epsilon}(\omega, k) P_{2m}(\omega, k) + R_4(\omega, k)
	\end{equation}
	of degree$~4$ in$~\omega$, where
		\begin{eqnarray}\label{2-poly-def}
		P_{2\epsilon}(\omega, k)= -\tau_{11} \omega^2 - i \omega + D_{\epsilon} k^2,\nonumber \\
		P_{2m}(\omega, k)= -\tau_{22} \omega^2 - i \omega + D_m k^2,
	\end{eqnarray}
	and the remainder
	\begin{eqnarray}
		R_4(\omega, k) = \left(  \bar{\kappa}_{11}c_m^{-1} -D_{\epsilon} \right) P_{2m}(\omega, k)  \nonumber \\ -\left(\omega^2 \tau_{12} - \bar{\kappa}_{12}\chi^{-1} k^2\right)
		 \left(\omega^2 \tau_{21} - \kappa_{21}c_m^{-1} k^2\right),
	\end{eqnarray}
	is a degree$~4$ polynomial in$~\omega$ of$~O(m_{z0}^2)$. The poles of Fourier-Laplace modes, or equivalently the zeros of$~P_4(\omega, k)$, lead to the dispersion relations.
	
	 In the absence of$~m_{z0}$ terms, energy and magnetization sectors decouple. In this case, the dispersion relations in the energy sector are obtained from the quadratic equation ${P_{2\epsilon}(\omega, k)=0}$, and are given by
	\begin{equation}\label{en-sec-DR}
		\omega_{\epsilon\pm}(k)= \left( -i \pm \sqrt{-1 + 4\tau_{11}D_{\epsilon}  k^2   } \right)/2\tau_{11}.
	\end{equation}
	When the time scale$~\tau_{11}$ associated with quasi-hydrodynamic energy current combines with diffusion coefficient$~D_{\epsilon}$ of hydrodynamic energy density, characteristic momentum scale$~k_1$ and velocity scale$~v_1$ emerge, given by the expressions
	\begin{equation}\label{k1v1-def}
		k_1=1/\sqrt{\tau_{11}D_{\epsilon}} ~, \quad v_1=\sqrt{D_{\epsilon}/\tau_{11}}.
	\end{equation}
	It follows from Eq.(\ref{en-sec-DR}) that the modes are diffusive when$~{|k| <  k_1/2}$ since the real part of$~\omega_{\epsilon\pm}(k)$ vanishes, while they propagate when$~{|k| > k_1/2}$, as plotted in~Fig.~\ref{fig1}. Essentially, in the regime$~{|k| \ll k_1}$, the two branches of dispersion relations reduce to
		\begin{equation}\label{en-sec-DR-small-1}
		\omega_{\epsilon+}(k)\approx -iD_{\epsilon} k^2  \left( 1 + \frac{k^2}{(k_1)^2}   \right),
	\end{equation}
	associated with the energy density mode, and 
			\begin{equation}\label{en-sec-DR-small-2}
		\omega_{\epsilon-}(k) \approx -\frac{i}{\tau_{11}}\left( 1-\frac{k^2}{(k_1)^2}   \right),
	\end{equation}
	associated with gapped current mode. Though the dispersion relations are modified due to presence of quasi-hydrodynamic modes in extended hydrodynamics, the nature of classical hydrodynamic modes are unchanged in this regime. While in the regime$~{|k| \gg k_1}$, the dispersion relations reduce to
		\begin{equation}\label{en-sec-DR-large}
	\omega_{\epsilon\pm}(k)\approx \pm v_1 |k| - \frac{i}{2\tau_{11}} ,
\end{equation}	
	resulting in entirely different sound mode propagating with velocity$~v_1$ and attenuating over a time$~\tau_{11}$. It is of course understood that any$~|k|$ is still smaller than the momentum scale$~\Lambda$ beyond which hydrodynamic description is invalid. Thus$~k_1$ is to be interpreted as the momentum scale where a diffusive and a gapped mode merge and metamorphose into two propagating modes. Note that while$~k_c=k_1/2$ is the actual point where the dispersion becomes non-analytic, we almost always identify it with$~k_1$ while referring to regions far away from it.
	
	In the magnetization sector, the dispersion relations$~\omega_{m\pm}(k)$ are similar to expressions in Eq.(\ref{en-sec-DR}) due to similarity between the two polynomials in Eq.(\ref{2-poly-def}). Expressions identical to those given in Eqs.(\ref{en-sec-DR-small-1}),(\ref{en-sec-DR-small-2}) and(\ref{en-sec-DR-large}) hold even in the magnetization sector, where the time scale$~\tau_{22}$ combined with diffusion coefficient$~D_m$ give rise to momentum scale$~k_2$ and velocity scale$~v_2$, given by the expressions
		\begin{equation}\label{k2v2-def}
		k_2=1/\sqrt{\tau_{22}D_m } ~, \quad v_2=\sqrt{D_m/\tau_{22}}.
	\end{equation}
	The parameter$~k_2$ can be interpreted as the momentum scale where the magnetization mode changes its nature from diffusive to propagating.
	
	In presence of$~m_{z0}$ terms, the dispersion relations in general cannot be obtained explicitly. In this case, apart from densities and currents being coupled, energy and magnetization sectors are also entwined making it harder to decipher the nature of ensuing modes. Nevertheless, we can perturbatively determine corrections to the dispersion relations specified in Eq.(\ref{en-sec-DR}) for small cross coefficients. The leading correction in the energy sector can be written as
	\begin{equation}\label{en-sec-DR-corr}
		\delta \omega_{\epsilon\pm}(k) = \frac{-R_4(\omega_{\epsilon\pm}(k), k) \left(\omega_{\epsilon\pm}(k) - \omega_{\epsilon\mp}(k)\right)^{-1}}{i \omega_{\epsilon\pm}(k) \left(\tau_{11} - \tau_{22}\right) +  \left( \frac{\tau_{22}}{\tau_{11} k_1^2} - \frac{\tau_{11}}{\tau_{22} k_2^2} \right) k^2 }.
	\end{equation}   
Similar corrections in magnetization sector can be obtained by taking$~{\epsilon \to m}$, ${\tau_{11} \leftrightarrow \tau_{22}}$ and ${k_1 \leftrightarrow k_2}$. The perturbative corrections modify the diffusion coefficients,  velocities of propagation and attenuation factors, but not the corresponding nature of the modes in both small and large regimes.

 To summarize, there are two momentum scales$~k_1$ and$~k_2$ in the paramagnetic phase which are associated with energy and magnetization sectors, respectively.  In the regime$~{|k| \ll k_1}~\text{and}~k_2$, the two classical modes are diffusive and the two extended ones are gapped. While in the other extreme, when$~{|k| \gg k_1}~\text{and}~k_2$, there are two sets of progressive modes. In the intermediate regimes,$~k_1 \ll |k| \ll k_2$ and $~k_2 \ll |k| \ll k_1$, whichever is applicable, there is one diffusive, one gapped, and a set of propagating modes. When$~k_1 < k_2$, the diffusive and gapped modes in the intermediate regime are associated with magnetization, while the propagating mode with energy. The association is reversed for the case$~k_1 > k_2$. Propagating modes in paramagnetic systems have been observed in experiments~\cite{MookLynn1985} and even analyzed in numerical studies~\cite{Tao2005physrevl}.

	\subsection{Dispersion relations in Ferromagnetic Phase}\label{ferro-modes}

In the ferromagnetic phase, the classical dispersion relations follow from the linearized equations that are obtained from Eqs.(\ref{con-eqs}),(\ref{phase-eqn}) and(\ref{cons-rel-2}) as follows:
\begin{eqnarray}\label{lin-ferro-cl}
		\frac{\partial \epsilon}{\partial t} &=& \kappa_{11}c_m^{-1}\nabla^2 \epsilon ,
	\nonumber \\ \frac{\partial m_z}{\partial t} &=&  \kappa_{22}\chi^{-1} \nabla^2 m_z + \rho_s\nabla^2 \theta, \nonumber\\
	\frac{\partial \theta}{\partial t} &=& {\chi}^{-1} m_z + \gamma \rho_s T^{-1} \nabla^2 \theta,
\end{eqnarray} 
where, we have used the linear relation$~{h_{\theta} = \rho_s v_{\theta} }$ with spin stiffness constant$~\rho_s$, and expanded about$~{m_z=0}, {\theta=0}$ and finite$~T$. The $\theta$-equation introduces another diffusion coefficient$~D_{\theta}$, and its coupling with$~m_z$-equation introduces a velocity scale$~v_{m\theta}$ or, equivalently, momentum scale$~k_{m\theta}$, given by the expressions
\begin{eqnarray}\label{DtVmKmt-def}
D_{\theta} = \gamma \rho_s / T  ,\quad
v_{m\theta} = \sqrt{\rho_s/\chi}  , \nonumber \\ k_{m\theta} = v_{m\theta}/\sqrt{D_{m} D_{\theta} } = \sqrt{T/(\gamma\kappa_{22}) }.
\end{eqnarray}
Other choices of momentum scale such as$~v_{m\theta}/D_{m}$ and$~v_{m\theta}/D_{\theta}$ are just equal to$~k_{m\theta}$ up to a factor that depends only on dimensionless parameter$~D_m/D_{\theta}$.
The existence of this momentum scale suggests the possibility that hydrodynamic modes with$~{|k| \ll k_{m\theta} }$ can be qualitatively different from those with$~{|k| \gg k_{m\theta} }$.

\begin{figure}[t]
	\centering
	\includegraphics[width=\columnwidth]{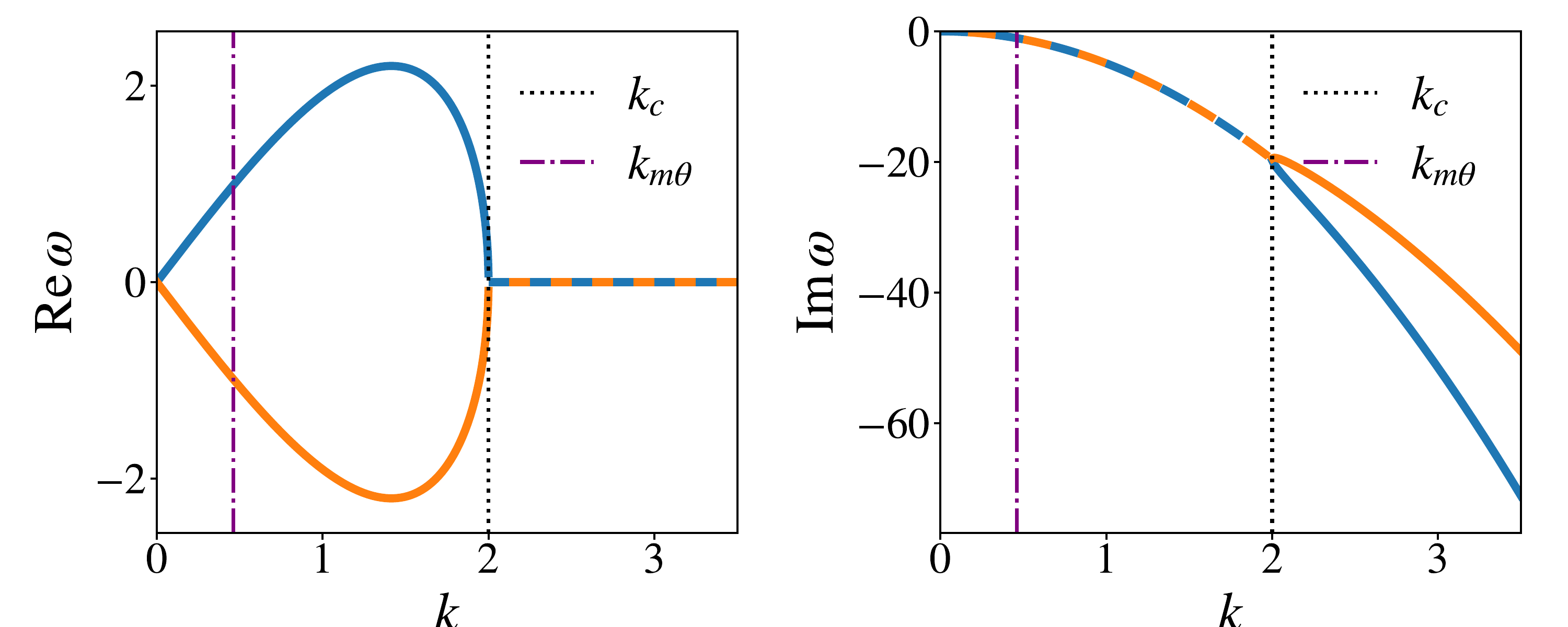}
	\caption{CIT dispersion relations of magnetization and phase sector in ferro-magnetic phase for the choice$~D_m=6, D_{\theta}=3.8$ and$~v_{m\theta}=2.2$. Cross-over from propagating to diffusive modes occurs at$~{k_c=2}$.}\label{fig2} 
\end{figure}

The energy density decouples from both$~m_z$ and$~\theta$, and follows the same dispersion as in paramagnetic phase,$~{\omega_{\epsilon} = -i D_{\epsilon} k^2 }$. From the Fourier-Laplace transformation of Eqs.(\ref{lin-ferro-cl}), we find that the other dispersion relations satisfy the polynomial equation
\begin{equation}\label{DR-pol-ferro-cl}
	\left(-i\omega + D_{m}k^2\right) 	\left(-i\omega + D_{\theta}k^2\right) + v_{m\theta}^2 k^2 =0,
\end{equation}
and are given by the expressions
\begin{equation}\label{DRs-ferro-cl}
	\omega_{m\theta\pm}= - \frac{i}{2} \Gamma_{m\theta} k^2 \pm \frac{i}{2}\sqrt{(D_{m}- D_{\theta})^2k^4 \!-\! 4 v_{m\theta}^2 k^2},
\end{equation}
where$~{\Gamma_{m\theta} = (D_{m} +  D_{\theta})}$. The modes are dispersive for$~{|k| > 2r k_{m\theta} }$, where ${r= 2/|\sqrt{D_m/D_{\theta}} - \sqrt{D_{\theta}/D_m}|}$, since the real part of$~\omega_{m\theta\pm}(k)$ vanishes, while they propagate for$~{|k| < 2r k_{m\theta} }$, as plotted in~Fig.~\ref{fig2}.
 Essentially, the dispersion relations become non-analytic at$~k_c=2r k_{m\theta}$. In other words, in the regime$~|k| \ll r k_{m\theta}$, the dispersion relations reduce to
\begin{equation}\label{omtpm}
		\omega_{m\theta\pm} \approx \pm v_{m\theta} |k| - \frac{i}{2} \Gamma_{m\theta} k^2,
\end{equation}
implying that the modes propagate with speed$~v_{m\theta}$ and damping rate$~\Gamma_{m\theta}$.
While in the regime$~|k| \gg r k_{m\theta}$, the modes decouple and follow diffusive dispersion given by the relations
\begin{equation}
	\omega_{m} \approx - i D_{m} k^2,~
	\omega_{\theta} \approx - i D_{\theta} k^2.
\end{equation}

We now proceed to establish the dispersion relations of extended hydrodynamics in the ferromagnetic phase. The linearized dynamical equations of the classical hydrodynamic variables, obtained from  Eqs.(\ref{con-eqs-ferro}) and(\ref{phase-eqn}), are
	\begin{eqnarray}\label{con-eqs-ferro-lin}
	&&\frac{\partial \epsilon}{\partial t} =   - \nabla \cdot J_{\epsilon}~,~ \frac{\partial m_z}{\partial t} = \rho_s\nabla^2 \theta  - \nabla \cdot J_m~,
	\nonumber\\
	&&\frac{\partial \theta}{\partial t} = {\chi}^{-1} m_z - \psi,
\end{eqnarray}
while the corresponding current equations, obtained from  Eqs.(\ref{curr-eqn-ferro-eit}), are
	\begin{eqnarray}\label{lin-ferro-eit}
	&&\tau_{11}\frac{\partial J_{\epsilon}}{\partial t} +  J_{\epsilon} = -\frac{\kappa_{11}}{c_m}\nabla \epsilon + \kappa_{11}T^2 \beta_{s1}\nabla \psi,
	\nonumber \\
	&&\tau_{22}\frac{\partial J_{m}}{\partial t} + J_{m} = - \frac{\kappa_{22}}{\chi} \nabla m_z, \nonumber \\
	&&\tau_{3}\frac{\partial \psi}{\partial t} + \psi = - \frac{\gamma \rho_s}{T} \nabla^2 \theta + \gamma \beta_{s1} \nabla \cdot J_{\epsilon}.
\end{eqnarray}
The dynamics is linearized about the steady state at$~{T\ne 0}, {h=0}, {h_{\theta}=0}$ and vanishing dissipative currents. The notation used for deviations from this state is same as that used for the variables. Here we have assumed due to their time-reversal properties that$~\beta_{s1}$ is a non-zero constant in general, while$~\beta_{s2}=0$. The time-reversal properties can be easily deduced from the fact that$~\epsilon$ and$~v_{\theta}$ are even under time reversal while$~m_z$ is odd, and their corresponding dissipative part of currents share the same properties. Hence, from Eq.(\ref{ent-flux-ferro}), it follows that$~\beta_{s1}$ is even variable while$~\beta_{s2}$ is odd and in turn of$~O(m_z)$. Note that unlike phase which only couples to magnetization in Eqs.(\ref{con-eqs-ferro-lin}), the phase current only couples to that of energy in Eqs.(\ref{lin-ferro-eit}). The transverse components of$~J_{\epsilon}$ and$~J_m$ decouple from the rest of the variables even in the ferromagnetic phase, and decay in time scales$~\tau_{11}$ and$~\tau_{22}$, respectively. The coupling of all three sectors, when$~{\beta_{s1} \ne 0}$, can lead to intricate behavior of the hydrodynamic modes.

The Fourier-Laplace transformation of  Eqs.(\ref{con-eqs-ferro-lin}) and(\ref{lin-ferro-eit}) are obtained as
	\begin{widetext}
	\begin{eqnarray}\label{ferro-eit-four}
		\begin{bmatrix}
			-i\omega & i|k| & 0 & 0 & 0 & 0\\
			i\kappa_{11} c_m^{-1}|k| & 1-i\omega \tau_{11} & 0 & 0 & 0 & -i\kappa_{11} T^2 \beta_{s1}|k|\\
			0 & 0 & -i\omega  &  i|k| & \rho_s k^2 & 0 \\
			0	& 0 & i \kappa_{22} \chi^{-1} |k|&  {1-i\omega \tau_{22} } & 0 & 0\\
			0	& 0 & -\chi^{-1} & 0 & -i\omega  & 1\\
			0	&-i \gamma \beta_{s1}|k|  & 0 & 0 & -\gamma \rho_s T^{-1} k^2  & 1-i\omega \tau_{3}
		\end{bmatrix}
		\begin{bmatrix}
			\tilde{\epsilon} (k,\omega)\\
			\tilde{J}^l_{\epsilon}(k,\omega)\\
			\tilde{m}_z(k,\omega)\\
			\tilde{J}^l_m(k,\omega)\\
			\tilde{\theta}(k,\omega)\\
			\tilde{\psi}(k,\omega)
		\end{bmatrix}
		=
		\begin{bmatrix}
			\epsilon (k, 0)\\
			\tau_{11}J^l_{\epsilon}(k,0)\\
			m_z(k,0)\\
			\tau_{22}J^l_m(k,0)\\
			\theta(k,0)\\
			\tau_3\psi(k,0)
		\end{bmatrix}.
	\end{eqnarray} 
\end{widetext}
We denote the determinant of the$~6\times 6$ matrix appearing in the above equation as$~P_6(\omega,k)$, which is a polynomial in$~\omega$ of at most degree$~6$. Solving the equation$~{P_6(\omega,k) =0}$ will lead up to six branches of dispersion relations, which in general cannot be determined explicitly.  

	\subsubsection{Decoupled Energy Sector}\label{z1-case}

Let us first consider the case$~\beta_{s1} = 0$ where the coupling between energy and phase currents is switched off. It is evident from the structure of the matrix that dispersion polynomial reduces to
\begin{equation}
	P^{(0)}_6(\omega,k) = 	P_{2\epsilon}(\omega,k)	P_{m\theta}(\omega,k),
\end{equation} 
where
\begin{eqnarray}
	&&P_{m\theta}(\omega,k) = 	P_{2m}(\omega,k)	P_{2\theta}(\omega,k) + v_{m\theta}^2 k^2 P_2(\omega),\nonumber \\
	&&P_{2\theta}(\omega,k) = -\tau_{3} \omega^2 - i \omega + D_{\theta} k^2,\nonumber \\
	&&P_2(\omega) = \left(1 - i \omega \tau_{22}\right) \left(1 - i \omega \tau_{3}\right).
\end{eqnarray}
The specific decomposition of the polynomial implies that the energy sector is decoupled containing in general two dispersion branches with behavior that is exactly same as in extended hydrodynamics of paramagnetic phase. Thus the corresponding modes with$~{|k| \ll k_1}$~(${|k| \gg k_1}$) are diffusive (propagating). The polynomial$~P_{m\theta}(\omega,k)$ associated with the coupled sectors of phase and magnetization has a suggestive extension of the expression appearing in Eq.(\ref{DR-pol-ferro-cl}), and in general contains four dispersion branches. It also clearly exhibits the existence of additional momentum scales. Apart from momentum scale$~k_2$ that we encountered in the paramagnetic phase,
the time scale$~\tau_{3}$ associated with the phase  current along with phase diffusion coefficient$~D_{\theta}$ result in a new momentum scale
\begin{equation}\label{k3-def}
	k_3=1/\sqrt{\tau_{3}D_{\theta}} .
\end{equation} It may appear that there is one more momentum scale
${\tilde{k}_{m\theta}=1/(v_{m\theta}\sqrt{\tau_{22}\tau_{3} })
}$,
 but is in fact not independent as it satisfies the relation$~{\tilde{k}_{m\theta}=k_2 k_3/k_{m\theta}}$.

\begin{figure}[t]
	\centering
	\includegraphics[width=\columnwidth]{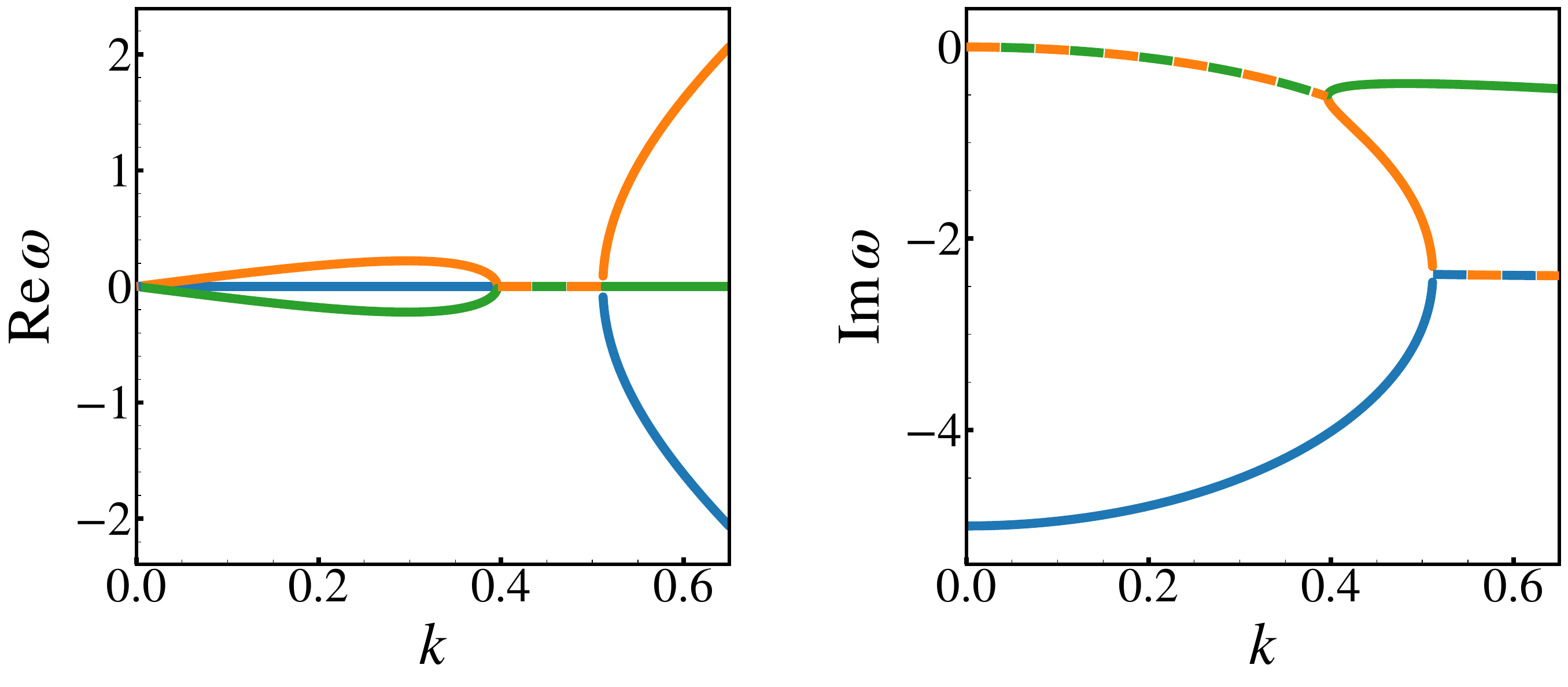}
	\caption{EIT dispersion relations for the case~${k_{m\theta} < k_2}$ in magnetization phase sector for the choice of parameters $D_m=5, D_{\theta}=0.5, v_{m\theta}=1.0, \tau_{22}=0.2,\tau_{11}=\tau_{3}=0$ and $\beta=0$.}\label{fig3a} 
\end{figure}

The limit$~k_2 \to \infty$ and$~k_3 \to \infty$ for a fixed$~k_{m\theta}$ amounts to completely decoupling the classical modes in the magnetization-phase sector from their extended ones. As we reduce$~k_2$ and$~k_3$ by increasing$~\tau_{22}$ and$~\tau_3$, respectively, while keeping CIT parameters$~D_m, D_{\theta}$ and$~v_{m\theta}$ fixed, we expect the deviations from classical hydrodynamics to become significant. Essentially, when a new momentum scale$~k_{new}$  approaches existing classical scale$~k_{old}$, new features are expected to be launched not only in the quasi-hydro range$~{|k| \gtrsim k_{new} }$, but also in the hydro-range$~{k_{new} \gtrsim |k| \gtrsim k_{old} }$. In fact, the dispersion relations in magnetization-phase sector as shown in~Fig.~\ref{fig2} get modified not only quantitatively but also qualitatively. Here, when either$~k_2$ or$~k_3$ is brought in as$~k_{new}$, we see that the modes in range$~{|k| \gtrsim k_{new} }$ are propagating and those in range$~{k_{new} \gtrsim |k| \gtrsim k_{m\theta} }$ transmute from diffusive to propagating ones.

\begin{figure}[t]
	\centering
	\includegraphics[width=\columnwidth]{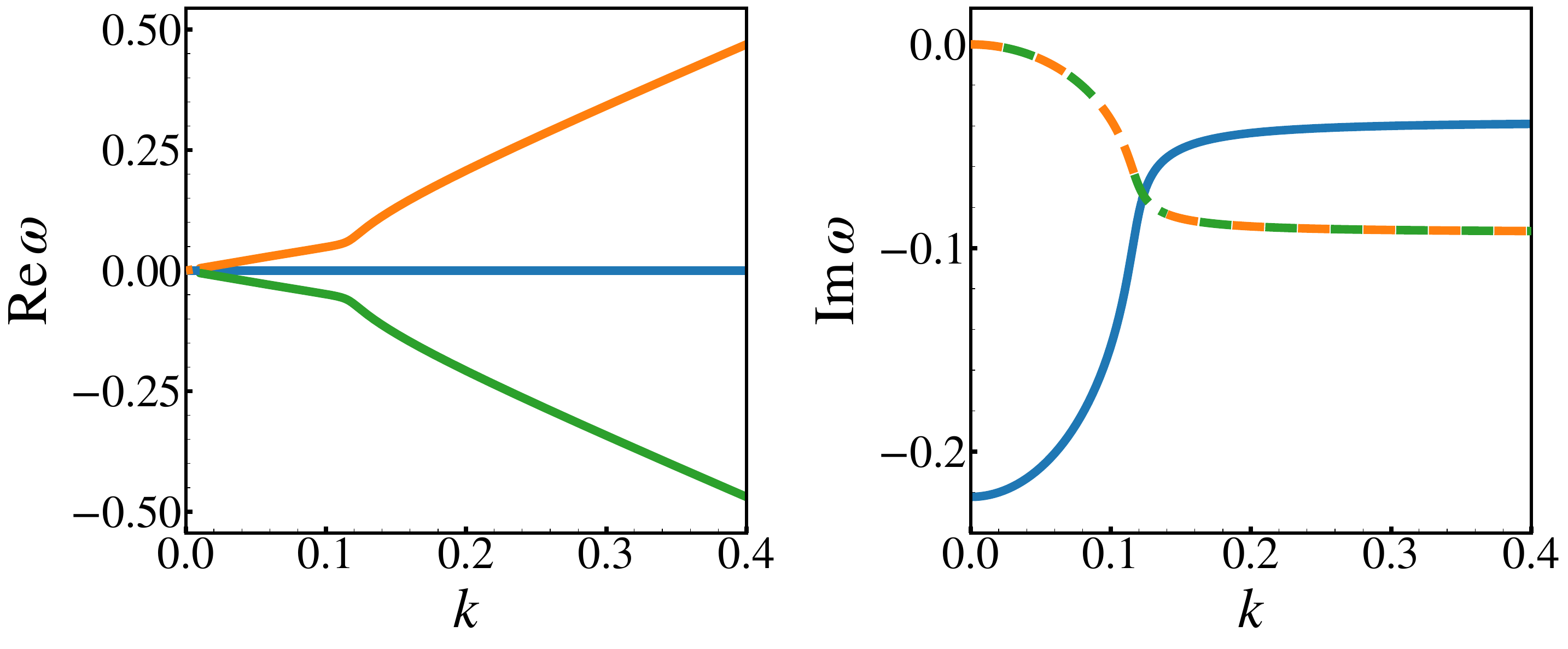}
	\caption{EIT dispersion relations for the case~${k_2 < k_{m\theta} }$ in magnetization phase sector for the choice of parameters $D_m=5.5, D_{\theta}=0.001, v_{m\theta}=0.5, \tau_{22}=4.5,\tau_{11}=\tau_{3}=0$ and $\beta=0$.}\label{fig3b} 
\end{figure}

Let us first tune in only one quasi-hydro scale, say$~k_2$, while keeping out$~{k_3 \to \infty}$, so as to simplify the classification of modes. In this case there are only three dispersion branches, and hence there can be at most one propagating mode. Here we can distinguish two cases: one where$~{k_{m\theta} < k_2}$ and the other where$~{k_2 < k_{m\theta}}$. Typical dispersion relations in (${k_{m\theta} < k_2}$)-case are shown in~Fig.~\ref{fig3a}. While these dispersion branches can exhibit intricate behavior, their general nature can be understood by zooming in regions that are far away from characteristic momentum scales and can be perturbatively analyzed. We find that, in regime ${ |k| \ll k_{m\theta} \ll k_2}$, there is a gapped mode$~{\omega=-i/\tau_{22} }$ and an attenuating sound mode$~\omega=\omega_{m\theta\pm}$ as given in Eq.(\ref{omtpm}) with velocity$~v_{m\theta}$; in regime ${ k_{m\theta} \ll |k| \ll k_2}$, the gapped mode continues to exist while the sound mode is replaced by two diffusive modes$~{\omega=-i D_m k^2}$ and$~{\omega=-i D_{\theta} k^2}$ associated with magnetization and phase; then in ${ k_{m\theta} \ll k_2 \ll |k|}$, phase diffusive mode$~{\omega=-i D_{\theta} k^2}$ continues to exist and a new sound mode ${\omega_{m\pm}(k) = \pm v_2 |k| - i/ (2\tau_{22})}$ with velocity$~v_2$ emerges,
similar to one we saw in energy sector given in Eq.(\ref{en-sec-DR-large}). These relations are obtained by solving cubic polynomial equation to the leading order in each regime; for instance, when ${ |k| \ll k_{m\theta} \ll k_2}$, we take ${ k^2 \sim \epsilon k^2_{m\theta} \sim \epsilon^2 k^2_2}$ and obtain dispersion relations perturbatively in$~\epsilon$. Similarly, typical dispersion relations in (${k_2 < k_{m\theta}}$)-case are shown in~Fig.~\ref{fig3b}. In this case, we find that, in the regime ${|k| \ll k_2 \ll k_{m\theta}}$, there is a gapped mode$~{\omega=-i/\tau_{22} }$ and a sound mode$~{\omega= \pm v_{m\theta} |k| }$; in the regime ${ k_2 \ll |k| \ll k_{m\theta}}$, the gapped mode continues to exist and the sound mode$~{\omega= \pm v_{m\theta} |k| -i D_{\theta}k^2/2}$ gets attenuated; while in the regime ${k_2 \ll k_{m\theta} \ll |k|}$, we see the phase difussive mode$~{\omega=-i D_{\theta} k^2}$ and a sound mode$~{\omega= \pm v_2 |k| -i v^2_{m\theta} /(2D_{\theta})}$ with velocity$~v_2$ and a modified attenuation factor. 

\begin{figure}[t]
	\centering
	\includegraphics[width=\columnwidth]{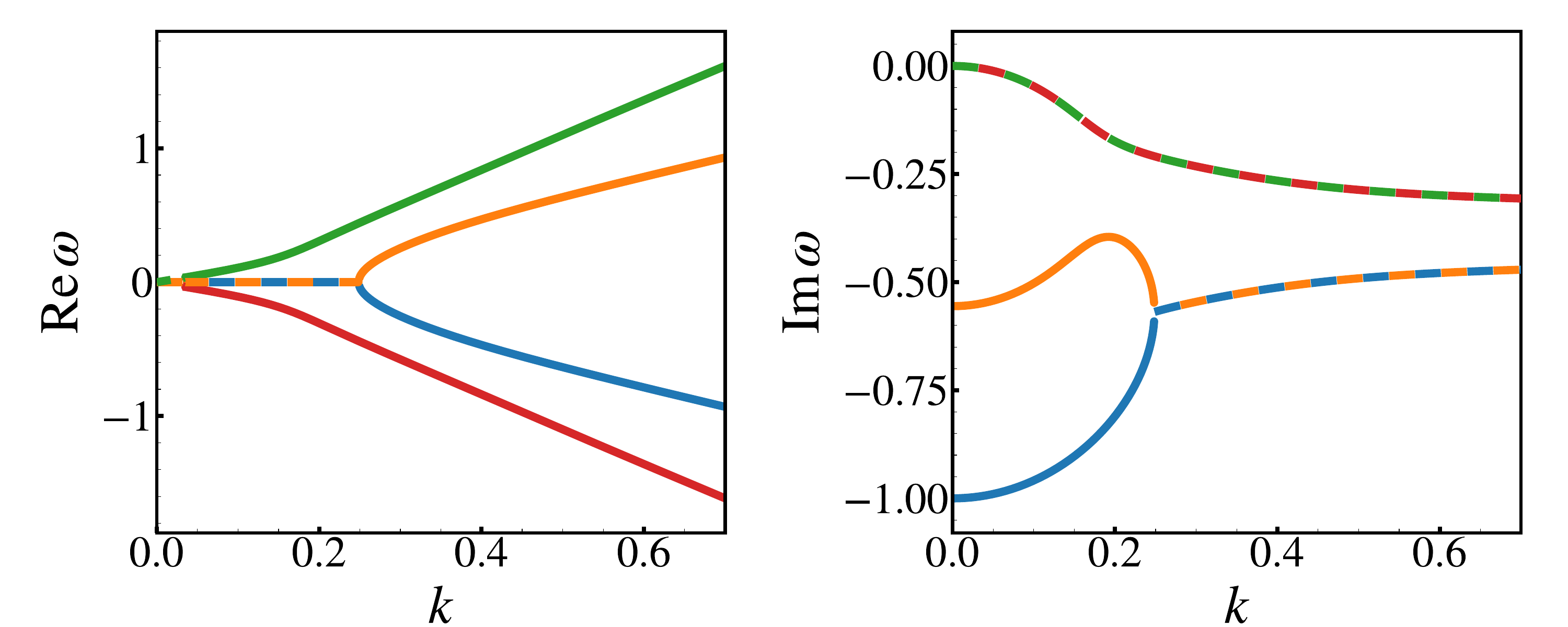}
	\caption{EIT dispersion relations for the case~${k_{m\theta} < k_2 < k_3}$ in magnetization phase sector for the choice of parameters $D_m=5, D_{\theta}=4, v_{m\theta}=1.0,\tau_{11}=0, \tau_{22}=1.8,\tau_{3}=1.0$ and $\beta=0$.}\label{fig3c} 
\end{figure}

If we now bring in the other scale$~k_3$, then$~P_{m\theta}(\omega,k)$ is a polynomial in$~\omega$ of degree 4, and hence there is an additional dispersion branch. This also leads to many distinguishable  cases:(i)$~{k_{m\theta} < k_2 < k_3}$; (ii)$~{k_2 < k_{m\theta} < k_3 }$; (iii)$~{k_2 < k_3  <  k_{m\theta} }$; and three other related cases obtained by interchanging$~k_2$ and$~k_3$ since$~P_{m\theta}(\omega,k)$ is invariant under the exchange of parameters ${D_m \leftrightarrow D_{\theta}}$ and ${\tau_{22} \leftrightarrow \tau_{3}}$. It is straightforward though cumbersome to perturbatively write down the dispersion relations in regions faraway from the three momentum scales, where we will encounter either two pairs of propagating modes, or one pair of propagating mode and two non-propagating modes, or else four non-propagating modes. Typical dispersion relations in ${k_{m\theta} < k_2 < k_3}$ case are shown in~Fig.~\ref{fig3c}. In this case, in regime$~{|k| \ll k_{m\theta} \ll k_2 \ll k_3 }$, there are two gapped modes$~\omega=-i/\tau_3$ and$~\omega=-i/\tau_{22}$, and sound mode$~{\omega=\omega_{m\theta\pm} }$ as given in Eq.(\ref{omtpm}); in regime$~{k_{m\theta} \ll |k| \ll k_2 \ll k_3 }$ the two gapped modes continue to exist while the sound mode is replaced by two diffusive modes$~\omega=-iD_m k^2$ and$~\omega=-iD_{\theta} k^2$; and then crosses over to two sound modes$~{\omega= \pm v_2 |k| - i/ (2\tau_{22})}$ and$~{\omega= \pm v_3 |k| - i/ (2\tau_{3})}$, in both $~{k_{m\theta} \ll k_2 \ll |k| \ll k_3 }$ and $~{k_{m\theta} \ll k_2 \ll k_3 \ll |k| }$ regimes. Note that for this specific choice of$~k_2$ and$~k_3$ in~Fig.~\ref{fig3c}, the two gapped branches collide and crosses over into sound, whereas for other choices diffusive branches can collide with their corresponding gapped branches to produce sound. 

	\subsubsection{Coupled Energy Sector}\label{z2-case}
\begin{figure}[t]
	\centering
	\includegraphics[width=\columnwidth]{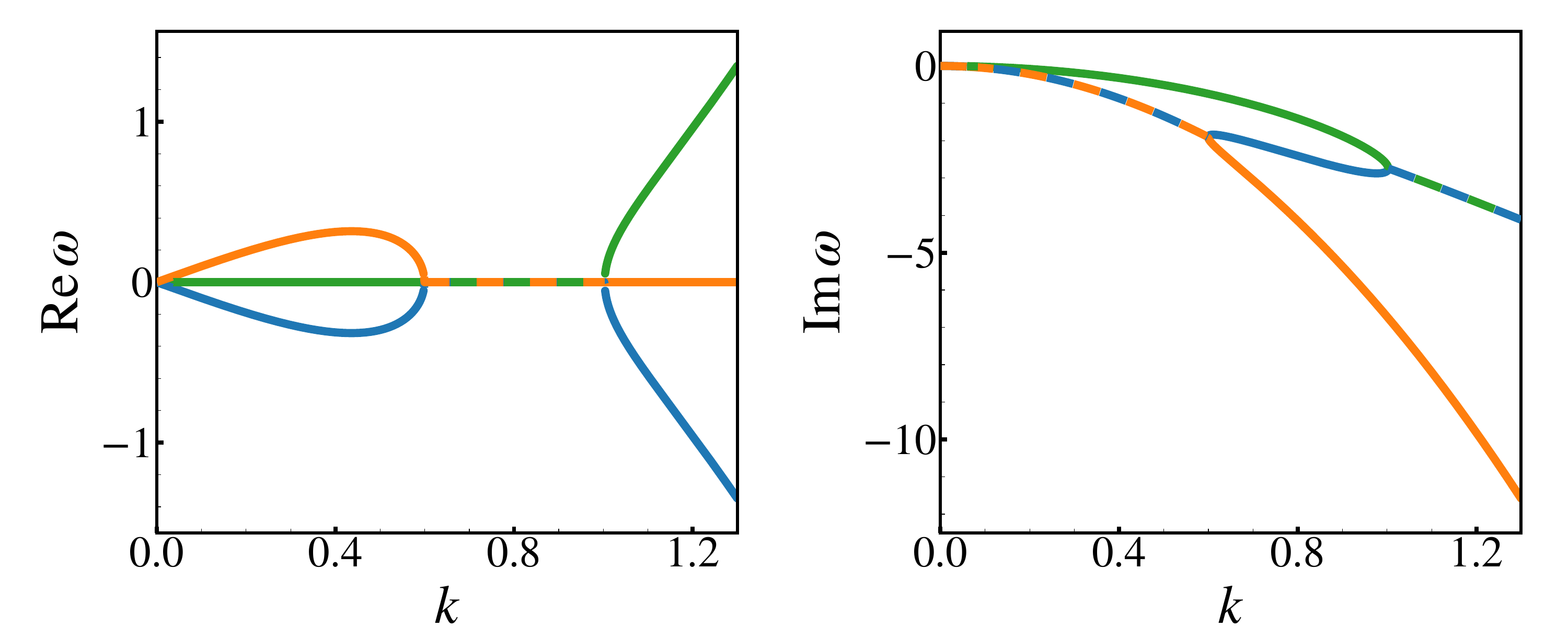}
	\caption{EIT dispersion relations for the case~${k_{m\theta} < k_{\beta}}$ for the choice of parameters $D_{\epsilon}=2, D_m=7, D_{\theta}=4, v_{m\theta}=1.0, \tau_{11}=\tau_{22}=\tau_{3}=0$ and $k_\beta=2.5$.}\label{fig3d} 
\end{figure}

When current coupling$~\beta_{s1} \ne 0$, the energy sector is no longer decoupled, and the dispersion polynomial can be written as
\begin{equation}\label{full-poly}
	P_6(\omega,k) = P^{(0)}_6(\omega,k) + i \omega 	P_3(\omega,k) \frac{k^2}{k_{\beta}^2},
\end{equation} 
where $k_{\beta} = 1/(T \beta_{s1}\sqrt{\gamma \kappa_{11}} )$, and 
\begin{equation}\label{P3-part}
	P_3(\omega,k) =\left(\omega^2 -v_{m\theta}^2 k^2 \right) \left(1 - i \omega \tau_{22}\right) + i \omega D_m k^2.
\end{equation}
The polynomial reveals at least three and at most six dispersion branches with plethora combination of propagating and non-propagating modes. The effect of current coupling in extended hydrodynamics is more revealing when we restrict the study to cases where either none of the relaxation time scales are switched on or only one of them is present. Typical dispersion relations when all relaxations times are switched off and with the choice$~{k_{m\theta} < k_{\beta}}$ are shown in~Fig.~\ref{fig3d}. This is the case that reveals the effect of current coupling$~\beta_{s1}$ on classical dispersion relations of energy, magnetization and phase. In the regime$~{|k| \ll k_{m\theta} \ll  k_{\beta}}$, classical behavior is unaffected leaving diffusive mode$~{\omega=-i D_{\epsilon}k^2}$ and sound mode$~{\omega=\omega_{m\theta\pm} }$ intact; in the regime$~{k_{m\theta} \ll |k| \ll k_{\beta}}$, all the three modes are diffusive with coefficients$~D_{\epsilon}, D_m$ and$~D_{\theta}$, respectively; and in regime$~{k_{m\theta} \ll k_{\beta} \ll |k|}$, the magnetization decouples and is diffusive with$~{\omega=-iD_m k^2}$, while energy and phase get coupled leading to sound mode$~{\omega= \pm v_{\epsilon\theta} |k| -i\Gamma_{\epsilon\theta}/2 }$, where$~{v_{\epsilon\theta}= k_{\beta}\sqrt{D_{\epsilon} D_{\theta}} }$ and$~{\Gamma_{\epsilon\theta} = (D_{\epsilon}+ D_{\theta}) k_{\beta}^2 }$. Essentially, this coupling drastically effects modes with large$~|k|$. If we instead choose$~{k_{\beta} < k_{m\theta}}$, we find that 
in small$~|k|$ region the behavior is similar to that in classical case; in the intermediate regime$~{k_{\beta}\ll |k| \ll k_{m\theta}}$, there is gapped mode$~{\omega= -i D_{\epsilon} k^2_{\beta} }$ and sound mode$~{\omega= \pm v_{m\theta} |k| -i D_m k^2/2}$ due to coupling of magnetization and phase; and in the large$~|k|$ region there is the diffusive mode of decoupled magnetization, and sound mode$~{\omega= \pm v_{\epsilon\theta} |k| -i v^2_{m\theta}/D_m }$ with a velocity that depends on coupling of energy and phase currents along with a modified attenuation factor that depends on the coupling of magnetization and phase. 

\begin{figure}[t]
  	\centering
  	\includegraphics[width=\columnwidth]{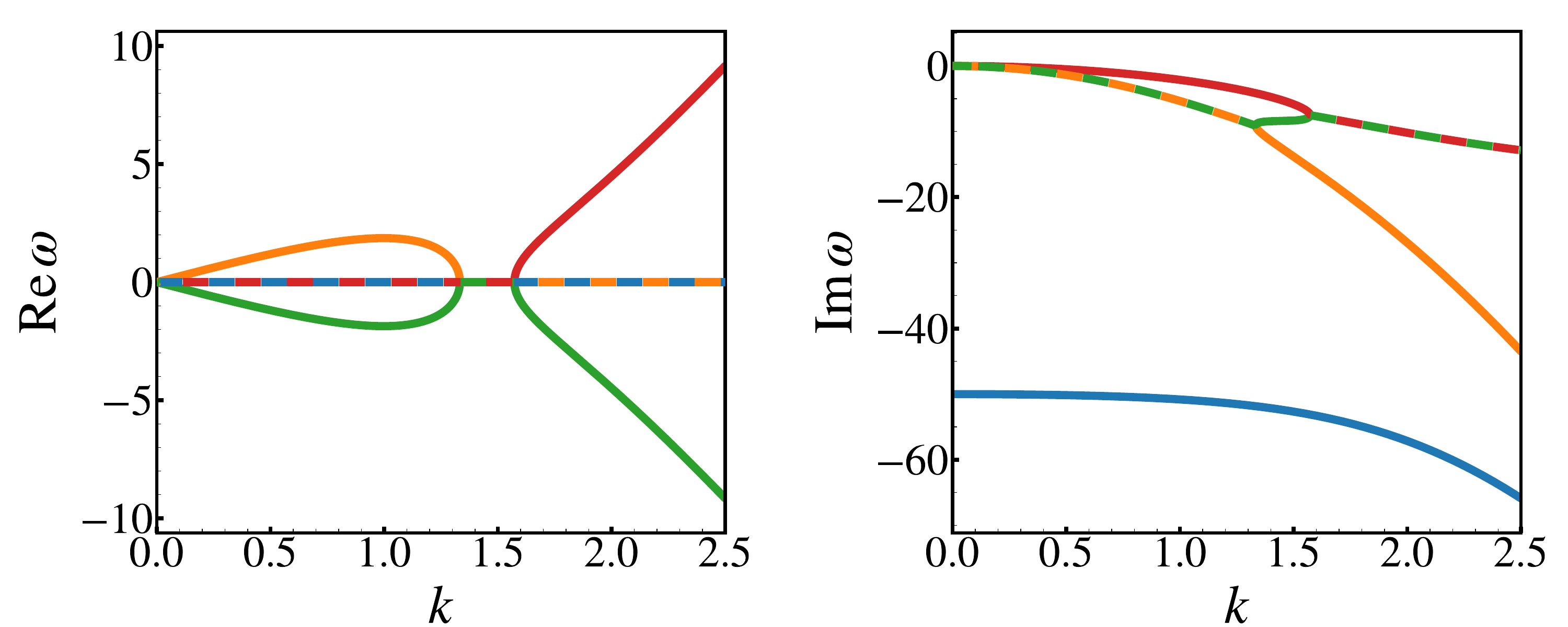}
  	\caption{EIT dispersion relations for the case~${k_{m\theta} < k_1 < k_{\beta}}$ for the choice of parameters $D_{\epsilon}=2, D_m=7.0, D_{\theta}=4, v_{m\theta}=2.5, \tau_{11}=0.02, \tau_{22}=\tau_{3}=0$ and $k_\beta=4.4$.} \label{fig3e} 
  \end{figure}

Let us now analyze the dispersion relation where only one relaxation time is switched on, and begin with the case where$~{\tau_{11} \ne 0}$ and$~{\tau_{22}=\tau_{3}=0}$. Typical dispersion relations
	with the choice$~{k_{m\theta} < k_1 < k_{\beta}}$ are shown in Fig.~\ref{fig3e}. In the small$~|k|$ regime$~{|k| \ll k_{m\theta} \ll k_1 \ll k_{\beta}}$, the energy sector decouples and contains its classical diffusive mode$~{\omega=-iD_{\epsilon}k^2}$ along with its extended gapped mode$~\omega=-i/\tau_{11}$, and sound mode$~{\omega=\omega_{m\theta\pm} }$ due to coupling of magnetization and phase. This sound mode gets replaced by diffusive modes with coefficients$~D_m$ and$~D_{\theta}$ in regime$~{k_{m\theta} \ll |k| \ll k_1 \ll k_{\beta}}$ leaving the other modes intact. In both$~{k_{m\theta} \ll k_1 \ll |k| \ll k_{\beta}}$ and$~{k_{m\theta} \ll k_1 \ll k_{\beta} \ll |k|}$ regimes, the energy modes consolidate into sound$~{\omega=\omega_{\epsilon\pm}(k)}$ as given in Eq.(\ref{en-sec-DR-large}). 

\begin{figure}[t]
		\centering
		\includegraphics[width=\columnwidth]{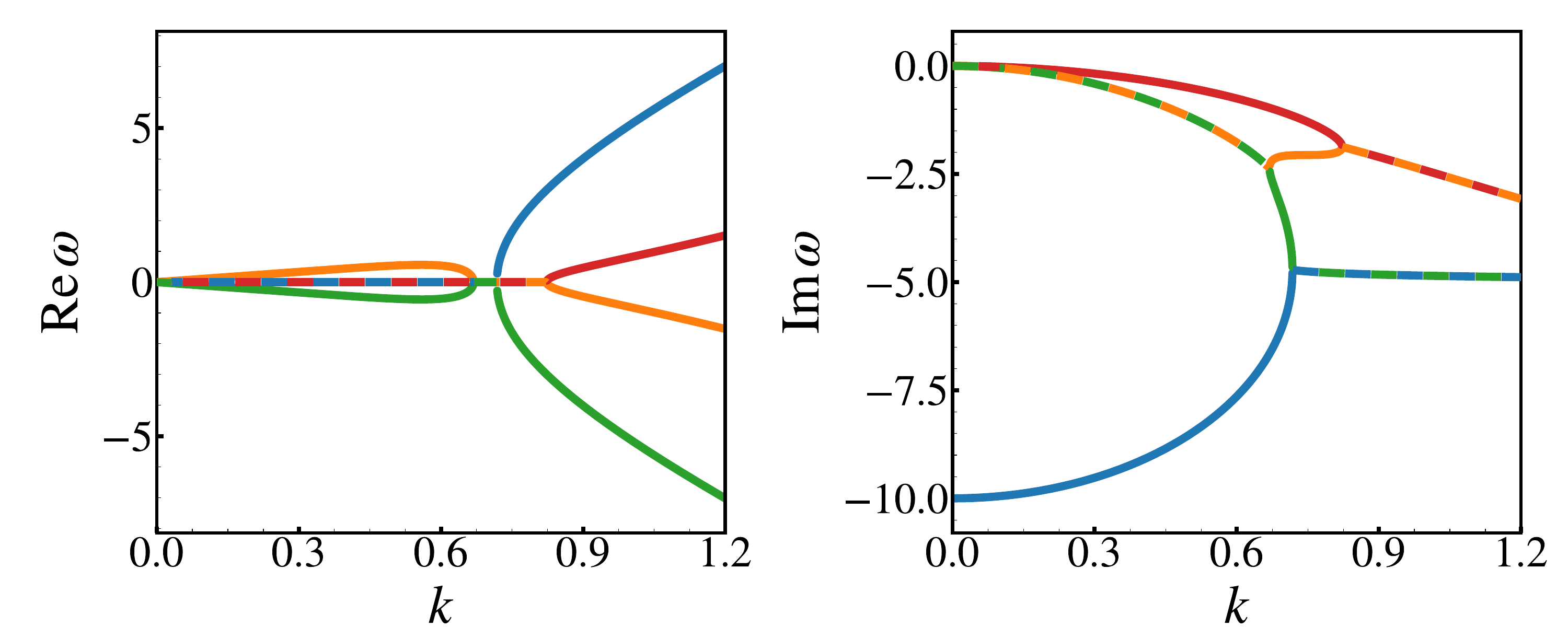}
		\caption{EIT dispersion relations for the case~${k_{m\theta} < k_2 < k_{\beta}}$ for the choice of parameters $D_{\epsilon}=2, D_m=5, D_{\theta}=4, v_{m\theta}=1.1, \tau_{22}=0.1, \tau_{11}=\tau_{3}=0$ and $k_\beta=1.8$.}\label{fig3f} 
	\end{figure}

   When relaxation time of magnetization current$~{\tau_{22}}$ is the only non-zero  time scale, typical dispersion relations are as shown in Fig.~\ref{fig3f}, for the choice$~{k_{m\theta} < k_2 < k_{\beta}}$. Apart from the addition of gapped mode$~\omega=-i/\tau_{22}$, in regimes$~{|k| \ll k_{m\theta} \ll k_2 \ll k_{\beta}}$ and $~{k_{m\theta} \ll |k| \ll k_2 \ll k_{\beta}}$, the classical hydro branches remain intact; in regime$~{k_{m\theta} \ll  k_2 \ll |k| \ll k_{\beta}}$, energy and phase stay diffusive while magnetization and its current make sound$~\omega=\pm v_2 -i/(2\tau_{22})$; while in regime$~{|k| \gg k_{\beta} }$ the diffusive modes are replaced by the sound$~{\omega= \pm v_{\epsilon\theta} |k| -i\Gamma_{\epsilon\theta}/2 }$ which is indicative of coupling between energy and phase currents. Note that by controlling magnetization current relaxation time we can turn diffusive energy mode into a propagating mode, which can be a desirable property if we do not want to heat up devices that are based on spin transport.  

 Finally, when we switch on only phase current relaxation time$~\tau_3$, the typical dispersion relations look similar to what is shown in~Fig.~\ref{fig3e} for$~\tau_{11} \ne 0$ case. In fact, for the choice$~{k_{m\theta} < k_3 < k_{\beta}}$, the behavior of modes in the first three intermediate regimes are more precisely as seen in$~\tau_{22} \ne 0$ case with interchange of parameters$~\{D_2,\tau_{22} \} \leftrightarrow \{D_3,\tau_{3} \}$. But the sound and two diffusive modes which appear in the third intermediate regime$~{k_{m\theta} \ll k_3 \ll |k| \ll k_{\beta}}$ continue to exist in$~{k_{m\theta} \ll k_3 \ll k_{\beta} \ll |k|}$ region too, unlike in$~\tau_{22} \ne 0$ case. Note that while$~P^{(0)}_6(\omega,k)$ in Eq.(\ref{full-poly}) preserves the symmetry under exchange of corresponding magnetization and phase parameters, the$~P_3(\omega,k)$ term which contains$~\tau_{22}$ breaks it. The effect of this symmetry breaking is the cause for distinct behavior of large$~k$ modes in$~\tau_{22} \ne 0$ case from that of$~\tau_3 \ne 0$ case.
 
 While the dispersion relations can be easily obtained by numerical analysis even when all the relaxation time scales are switched on, they may not prove revealing in these cases due to their complexity. But interestingly, this analysis suggests that we can tune these parameters to engineer a desired dispersion relation.  

	%
	%
	
	\section{Conclusion}\label{conc} 
	
	Recent experimental advances are enabling us to probe hydrodynamic regime and beyond in a variety of systems including spin systems. Understanding results that are at variance with classical hydrodynamics requires identification of relevant quasi-hydro variables and including them in the dynamical description. An established framework wherein the relevant variables are identified as dissipative currents and included in an extended hydrodynamics is based on principles of extended irreversible thermodynamics. We have employed this framework in explicitly constructing the extended hydrodynamics of planar Heisenberg ferromagnets in both paramagnetic and ferromagnetic phases.
	
	In the paramagnetic phase, energy and magnetization dissipative currents become dynamical and the extended hydrodynamics contains four relaxation time parameters$~\{\tau_{ij}\}$. The diagonal relaxation time scales$~\{\tau_{11},\tau_{22} \}$ survive the linearization of extended hydrodynamics, and along with diffusive coefficients$~\{D_{\epsilon}, D_m\}$ of classical hydrodynamics provide not only new momentum scales$~\{k_1, k_2\}$ and velocity scales$~\{v_1, v_2\}$ but also modified dispersion relations and new physics. When linearized about equilibrium states with finite temperature and zero magnetization, energy and magnetization sectors decouple, each containing a sound mode at large$~|k|$ contrary to what classical hydrodynamics would predict.
	
	 In the ferromagnetic phase, the phase current also becomes dynamical and its extended hydrodynamics contains additional relaxation time parameter$~\tau_{3}$ and current coupling parameters$~\beta_{s1}$ and$~\beta_{s2}$. In the linearized extended hydrodynamics$~\tau_{3}$ and$~\beta_{s1}$ survive along with diffusion coefficient$~D_{\theta}$ and coupling$~v_{m\theta}$ of classical hydrodynamics. The three diffusion coefficients, three relaxation time scales and two couplings result in five characteristic momentum scales$~\{k_{m\theta}, k_1, k_2, k_3, k_{\beta}\}$ which can be ordered in 120 different ways offering plethora combination of dispersion relations. In the limit$~{ k_{\beta} \to \infty }$, the energy sector decouples carrying away the scale$~k_1$, while the combined magnetization and phase sector is left with the scales$~\{k_{m\theta}, k_2, k_3\}$ along with the symmetry under exchange$~\{D_2,\tau_{22} \} \leftrightarrow \{D_3,\tau_{3} \}$. The dispersion relations of the energy sector are similar to paramagnetic case and contains sound mode at large$~|k|$. The scales in magnetization and phase sector can be ordered in six different ways tuning each scale as desired. We have presented typical dispersion relations for cases$~{(k_{m\theta}< k_2)}$ and$~{(k_2 < k_{m\theta})}$, in the limit$~{k_3 \to \infty}$, and for the$~{(k_{m\theta}< k_2 < k_3)}$, and explicitly obtained analytical expressions in the intermediate regimes where$~|k|$ is far away from all the momentum scales. We further considered cases with at most one of the relaxation time scales and finite$~k_{\beta}$. These cases illustrate a rich display of dispersion relations containing multitude of combinations of non-propagating gapped and diffusive modes, and propagating sound modes with and without attenuation. For instance, when$~{k_{\beta} \ne 0}$, we have seen that diffusive energy mode could be changed to propagating one, if we so desire, by either tuning on$~\tau_{11}$ or$~\tau_{22}$. While it is not surprising to make this change by controlling$~\tau_{11}$, it is interesting to see that it can also be done by controlling relaxation time scale of magnetization current.

	While we have not analyzed the dispersion relations of every conceivable arrangement of scales, our analysis illustrates that a myriad of combinations of hydro modes can be engineered by tuning in quasi-hydro transport parameters. We also demonstrated that the modes undergo non-analytic changes in nature controlled by the characteristic momentum scales. Further, we have described how to explicitly find the nature of modes when$~|k|$ is far away from all the characteristic momentum scales. The formulation developed here to analyze the planar Heisenberg model can easily be applied in effecting extended hydrodynamics of other Heisenberg spin models, itinerant spin systems, anti-ferromagnets and superfluids, to name a few. It is useful both in incorporating quasi-hydro variables and their interactions beyond local equilibrium, and analyzing their effects on non-equilibrium transport in a wide variety of macroscopic systems.

	%
	%

\bibliographystyle{apsrev4-2}
\bibliography{references}

\end{document}